\documentclass[preprint]{elsarticle}

\usepackage{euscript,amssymb,amsmath}
\usepackage{amsfonts,latexsym}
\usepackage{amsfonts,amsmath,amssymb,amsfonts,latexsym}
\usepackage[utf8]{inputenc}

\usepackage{graphicx}
\usepackage{epsfig}
\usepackage{color}

\newcommand{\be}[1]{\begin{equation}\label{#1}}
\newcommand{\ee}{\end{equation}}
\newcommand{\ba}[1]{\begin{eqnarray}\label{#1}}
\newcommand{\ea}{\end{eqnarray}}
\newcommand{\rf}[1]{(\ref{#1})}
\newcommand{\nn}{\nonumber}

\newcommand{\sfrac}[2]{\dfrac{\,#1\,}{\,#2\,}}

\def\H{{\cal H}}      

\begin{document}
	
\begin{frontmatter}
		
\title{Backreaction in cosmic screening approach}

\author[a]{Maxim Eingorn}
\ead{maxim.eingorn@gmail.com}

\author[a,b]{Brianna O'Briant}
\ead{bsobrian@ncsu.edu}

\author[a]{Adjaratou Diouf}
\ead{adiouf1@eagles.nccu.edu}
		
\author[c,d]{Alexander Zhuk}
\ead{ai.zhuk2@gmail.com}

\address[a]{Department of Mathematics and Physics, North Carolina Central University,\\ 1801 Fayetteville St., Durham, North Carolina 27707, U.S.A.}

\address[b]{Department of Mechanical and Aerospace Engineering, North Carolina State University,\\ Campus Box 7910, Raleigh, North Carolina 27695, U.S.A.}

\address[c]{Astronomical Observatory, Odessa National University,\\ Dvoryanskaya St. 2, Odessa 65082, Ukraine}
		
\address[d]{Center for Advanced Systems Understanding (CASUS),\\
Untermarkt 20, 02826 G\"{o}rlitz, Germany}

\begin{abstract} We investigate the backreaction of  nonlinear perturbations on the global evolution of the Universe within the cosmic screening approach. To this end, we have considered the second-order scalar perturbations. An analytical study of these perturbations followed by a numerical evaluation shows that, first,  the corresponding average values  have a negligible backreaction effect on the Friedmann equations and, second, the second-order correction to the gravitational potential is much less than the first-order quantity. Consequently, the expansion of perturbations into orders of smallness in the cosmic screening approach is correct. \end{abstract}

		
\end{frontmatter}
	
\section{Introduction}
	
\setcounter{equation}{0}

In accordance with the Cosmological Principle, there should not be selected directions and positions in the Universe. That is, a sufficiently large volume, chosen arbitrarily in the Universe, should contain approximately the same amount of matter. From a physical point of view, this is a fairly reasonable assumption, which, in general, is confirmed by observations. Although it should be noted that the scale of homogeneity (also called the cell of uniformity) is the subject of debate (see the corresponding references e.g. in \cite{Garc}, \cite{Andrade}). For example, analysis of the quasar sample of the Sloan Digital Sky Survey indicates the scale of homogeneity of the order of $70\div 90 h^{-1}$ Mpc \cite{Goncalves}, while N-body simulations of the $\Lambda$CDM  model estimate this scale to be about $260 h^{-1}\approx 370$ Mpc for $h=0.7$ \cite{Yadav}. Moreover, the observations indicate the existence of such super-large objects as the Sloan Great Wall $\sim 420$ Mpc \cite{27}, the Huge LQG $\sim 1.2$ Gpc \cite{30},  the Giant GRB ring $\sim 1.7$ Gpc \cite{31},  and the Hercules-Corona Borealis Great Wall $\sim 2\div 3$ Gpc \cite{32}. Recently, the Giant Arc $\sim 1$ Gpc \cite{33} was also found. Therefore, the scale of homogeneity can be greatly increased. It is worth noting that the size of the  Hercules-Corona Borealis Great Wall is of the order of the characteristic screening length $\lambda_{\mathrm{eff}}$ of the gravitational interaction in the Universe \cite{EE}. According to the cosmic screening approach \cite{Ein1,EB,EKZ1,EKZ2},  the screening length defines the upper bound for the dimensions of a solitary structure in the Universe. Therefore, structures larger than Hercules-Corona Borealis Great Wall should not exist. 

At scales larger than the scale of homogeneity, the averaged evolution of the Universe  is governed by the
Friedmann-Lema$\mathrm{\hat{\i}}$tre-Robertson-Walker (FLRW) metric
\ba{1.1}
ds^2 = a^2(\eta) \big[ d\eta^2 -  \,\delta_{\alpha\beta}dx^{\alpha}dx^{\beta} \big]\, , \ea
where the scale factor $a$ depends only on the conformal time $\eta$, and we consider a spatially flat model. 
In the case of the $\Lambda$CDM model, the Friedmann equations read
\be{1.2}
\sfrac{3\H^2}{a^2} = \kappa \bar{\varepsilon} + \Lambda
\ee
and
\be{1.3}
\frac{2\H'+\H^2}{a^2}=\Lambda\, ,
\ee
where $\H\equiv a'/a \equiv (da/d\eta)/a$ and the average energy density of the nonrelativistic pressureless ($\overline{p}=0$) matter \mbox{$\bar\varepsilon = \bar{\rho}c^2/a^3$}, with $\bar\rho$ being the average comoving mass density. We also define $\kappa \equiv 8\pi G_{\!N}/c^4$, where $G_{\!N}$ and $c$ denote the gravitational constant and the speed of light. 

Small-scale inhomogeneities, e.g. galaxies and groups of galaxies, perturb the averaged metric \cite{Mukhanov,Durrer,Rubakov}:
\ba{1.4}
ds^2 = a^2(\eta) \left[ \left(1+2\Phi +2\Phi^{(2)}\right)d\eta^2 - \left(1-2\Psi -2\Psi^{(2)}\right) \,\delta_{\alpha\beta}dx^{\alpha}dx^{\beta} \right]\, , \ea
where for the purpose of our paper we consider only scalar perturbations. Here, $\Phi, \Psi$ and $\Phi^{(2)},\Psi^{(2)}$ are the scalar perturbations of the first and second order, respectively. In the case of $\Lambda$CDM model, the first-order perturbations $\Phi$ and $\Psi$ are equal: $\Phi=\Psi$, and their average values should be equal to zero: $\overline{\Phi}=0$. On the other hand, average values of the second-order perturbations $\overline{\Phi^{(2)}}$, $\overline{\Psi^{(2)}}$, and also such quantities as $\overline{\Phi^2}$, $\overline{\rho\Phi}$, etc. in general are not equal to zero and have a backreaction effect on the global evolution of the Universe. This is the backreaction problem (see, e.g., the reviews \cite{BR1,BR2,BR3,BR4,BR5,BR6,BR7} and references therein). 

The purpose of this article is to investigate how strongly the perturbations affect the averaged behavior of the Universe. If such an influence is significant, then, firstly, the expansion of perturbations in terms of the degree of smallness, as done in \rf{1.4}, is not correct, and, secondly, the Friedmann equations, for instance \rf{1.2} and \rf{1.3}, must be modified. To address these questions,
we consider the relativistic theory of perturbations within the cosmic screening approach \cite{Ein1,EB,EKZ1,EKZ2}. This approach got its name due to the fact that the gravitational potential created by all inhomogeneities satisfies the Helmholtz equation, and not the Poisson one. Therefore, individual inhomogeneities make contributions in the form of the Yukawa potential with the characteristic screening length \cite{EE,Ein1}. In this approach, there is a clear division of perturbations according to orders of smallness. The first-order quantities contribute to the sources of the second-order ones. As a result, all equations are linear. This made it possible to solve them analytically in the case of the $\Lambda$CDM model 
(see \cite{EE,Ein1} and \cite{second-order1,second-order2} for the first and second orders, respectively). This is an important point, since it allows us to study the backreaction problem also analytically. As we show in our article, the velocity-independent nonlinear quantities of the form $\overline{\Phi_0^2}$ and $\overline{\rho\Phi_0}/\overline{\rho}$ do not have a significant backreaction effect on the Friedmann equation, and the second-order value of ${\Psi_0^{(2)}}$ is much smaller than the first-order values.

The paper is structured as follows. In Section 2, we study the backreaction effect of the second-order scalar perturbations on the Friedmann equations within the cosmic screening approach. In Section 3, we demonstrate that the second-order correction to the gravitational potential is much less than the first-order quantity. The main results are summarized in concluding Section 4.

\section{Backreaction in Friedmann equations}

Within the cosmic screening approach, the perturbed Friedmann equations can be obtained with the help of the Euclidean averaging \cite{BR4} of the equations (3.36) and (3.38) in \cite{second-order1}:
\be{2.5}
\sfrac{3\H^2}{a^2} - \frac{6\H}{a^2}\left[ \H \overline{\Phi^{(2)}} +\overline{(\Psi^{(2)})'}\, \, \right]  - 3\kappa\overline{\varepsilon}\overline{\Psi^{(2)}}= \kappa \bar{\varepsilon} + \Lambda +\kappa\overline{\varepsilon}^{(\mathrm{II})}
\ee
and
\ba{2.6}
\frac{2\H'+\H^2}{a^2} 
- \frac{2}{a^2}\left[ \, \,  \overline{(\Psi^{(2)})''} +2\H\overline{(\Psi^{(2)})'} + \H\overline{(\Phi^{(2)})'} + (2\H'+\H^2)\overline{\Phi^{(2)}}\, \right]  \nn\\
=\Lambda- \kappa\overline{p}^{(\mathrm{II})}\, ,
\ea
where, up to the fourth-order correction terms, the effective average energy density $\overline{\varepsilon}^{(\mathrm{II})}(\eta)$ and pressure $\overline{p}^{(\mathrm{II})}(\eta)$ read (see Eqs. (4.5) and (4.6) in \cite{second-order1}):
\ba{2.7}
\kappa\overline{\varepsilon}^{(\mathrm{II})} = \frac{\kappa c^2}{2a^3}\overline{\rho\Phi_0}-\frac{15}{a^2}\H^2\overline{\Phi^2_0} = \frac{\kappa}{2}\overline{\varepsilon}\frac{\overline{\rho\Phi_0}}{\overline{\rho}} - 5\left( \kappa\overline{\varepsilon}+\Lambda\right) \overline{\Phi^2_0}
\ea
and
\ba{2.8}
\kappa\overline{p}^{(\mathrm{II})}=
\frac{\kappa c^2}{6a^3}\overline{\rho\Phi_0}-\left( \frac{7\kappa\overline{\rho}c^2}{2a^3}- \frac{5}{a^2}\H^2\right) \overline{\Phi^2_0} =
\frac{\kappa}{6}\overline{\varepsilon}\frac{\overline{\rho\Phi_0}}{\overline{\rho}} - \left(\frac{11}{6} \kappa\overline{\varepsilon}-\frac{5}{3}\Lambda\right) \overline{\Phi^2_0}\, .
\ea
Here, we consider only velocity-independent contributions in connection with a toy model analyzed below. In particular, $\Phi_0$ is the velocity-independent part of the first-order scalar perturbation $\Phi$ \cite{Ein1}:
\be{2.9}
\Phi_0=\frac{1}{3}-\frac{\kappa c^2}{8\pi a}\sum_n \frac{m_n}{|{\bf r}-{\bf r}_n|}\exp\left(-\frac{a|{\bf r}-{\bf r}_n|}{\lambda}\right)\, .
\ee
We consider a model where inhomogeneities are presented in the form of a system of separate nonrelativistic point-like particles with masses $m_n$ and comoving radius-vectors ${\bf r}_n(\eta)$. The screening length $\lambda$ is
\be{2.10}
\lambda = \sqrt{\frac{2 a^3}{3\kappa \overline{\rho}c^2}}\, .
\ee
The average value of $\Phi_0$ is equal to zero as it should be for the first-order perturbation.

Eqs. \rf{2.5} and \rf{2.6} demonstrate that the effective average energy density $\overline{\varepsilon}^{(\mathrm{II})}$ and pressure $\overline{p}^{(\mathrm{II})}$,  on the one hand,  serve as sources of the second-order perturbations and, on the other hand, they provide the backreaction to the background Friedmann equations. This backreaction is significant if these quantities are of the order of the background  parameters $\overline{\varepsilon}$ and $\Lambda$. This will be the case if the quantities $\overline{\rho\Phi_0}/\overline{\rho}, \overline{\Phi^2_0}\sim o(1)$, as it follows from Eqs. \rf{2.7} and \rf{2.8}. Thus, the task is to estimate these quantities. 

According to Eqs. (4.9) and (4.10) in \cite{second-order1},
\be{2.11} 
\overline{\rho\Phi_0} = \frac{1}{3}\overline{\rho}-\frac{\kappa c^2}{8\pi a}\,\frac{1}{\mathcal{V}}\sum\limits_{n}\sum\limits_{k\neq n}\frac{m_n
	m_k}{|{\bf r}_k-{\bf r}_n|}\exp\left(-\frac{a|{\bf r}_k-{\bf r}_n|}{\lambda}\right)\, \ee
and 
\be{2.12} 
\overline{\Phi_0^2} = -\frac{1}{9} +\frac{\kappa c^2}{48\pi\overline{\rho}\lambda}\,\frac{1}{\mathcal{V}} \sum\limits_{n}\sum\limits_{k}m_n m_k
\exp\left(-\frac{a|{\bf r}_k-{\bf r}_n|}{\lambda}\right)\, ,\ee
where $\mathcal{V}$ is an averaging volume. In order to perform further evaluations, we consider a toy model in which all particles (galaxies) have the same masses and are located at the same distance $l$ from each other (like a crystal lattice with a period equal to the average distance between galaxies). Gravitational potentials and forces within this model have been previously analyzed in \cite{TTT1,TTT2}. Let us choose some particle as the origin of the reference frame ${\bf r}_n=0$ and sum over all other particles $\sum\limits_{k}$. It is clear that for each $n$-th particle we obtain the same result. Then, we get
\be{2.13} 
\overline{\rho\Phi_0} = \frac{1}{3}\overline{\rho}-\frac{\kappa c^2}{8\pi a}\,\frac{1}{\mathcal {V}}Nm^2\sum\limits_{k}\frac{1}{ r_k}\exp\left(-\frac{ar_k}{\lambda}\right),\quad r_k\neq 0\, ,
\ee
\be{2.14} 
\overline{\Phi_0^2} = -\frac{1}{9} +\frac{\kappa c^2}{48\pi\overline{\rho}\lambda}\,\frac{1}{\mathcal{V}} Nm^{2}\sum\limits_{k}
\exp\left(-\frac{ar_k}{\lambda}\right)\, ,
\ee
where $N$ is the number of particles in the volume $\mathcal{V}$.
Replacing $\sum\limits_k$ by the triple sum $\sum\limits_{k_1=-\infty}^{+\infty}\sum\limits_{k_2=-\infty}^{+\infty} \sum\limits_{k_3=-\infty}^{+\infty}$ and $r_k$ by $l\sqrt{k_1^2+k_2^2+k_3^2}$, and taking into account that $Nm/\mathcal{V}=\overline\rho$ and $m=\overline\rho l^3$ (since,
in the toy model under consideration, there is one mass $m$ in the comoving volume $l^3$), instead of \rf{2.13} and \rf{2.14} we get, respectively,
\ba{2.15} 
\overline{\rho\Phi_0} = \frac{1}{3}\overline{\rho}-\frac{\kappa c^2\overline\rho^2l^2}{8\pi a}\sum\limits_{k_1=-\infty}^{+\infty}\sum\limits_{k_2=-\infty}^{+\infty}\sum\limits_{k_3=-\infty}^{+\infty}\frac{1}{ \sqrt{k^2_1+k^2_2+k^2_3}}\nn\\
\times\exp\left(-\frac{al\sqrt{k^2_1+k^2_2+k^2_3}}{\lambda}\right)\, ,
\ea
where $k_1^2+k_2^2+k_3^2\neq0$, and
\be{2.16} 
\overline{\Phi_0^2} = -\frac{1}{9} +\frac{\kappa c^2\overline\rho l^3}{48\pi\lambda}\sum\limits_{k_1=-\infty}^{+\infty}\sum\limits_{k_2=-\infty}^{+\infty}\sum\limits_{k_3=-\infty}^{+\infty}
\exp\left(-\frac{al\sqrt{k^2_1+k^2_2+k^2_3}}{\lambda}\right)\, .
\ee
Finally, introducing $\tilde\lambda\equiv\lambda/(al)$, we obtain
\ba{2.17} 
\overline{\rho\Phi_0} = \frac{1}{3}\overline{\rho}\left[1-\frac{1}{4\pi \tilde\lambda^2}\sum\limits_{k_1=-\infty}^{+\infty}\sum\limits_{k_2=-\infty}^{+\infty}\sum\limits_{k_3=-\infty}^{+\infty}\frac{1}{ \sqrt{k^2_1+k^2_2+k^2_3}}\right.\nn\\
\left.\times\exp\left(-\frac{\sqrt{k^2_1+k^2_2+k^2_3}}{\tilde\lambda}\right)\right]\, ,
\ea
where $k_1^2+k_2^2+k_3^2\neq0$, and
\be{2.18} 
\overline{\Phi_0^2} = -\frac{1}{9}\left[1-\frac{1}{8\pi\tilde\lambda^{3}}\sum\limits_{k_1=-\infty}^{+\infty}\sum\limits_{k_2=-\infty}^{+\infty}\sum\limits_{k_3=-\infty}^{+\infty}
\exp\left(-\frac{\sqrt{k^2_1+k^2_2+k^2_3}}{\tilde\lambda}\right)\right]\, .
\ee

\begin{figure*}[!ht]
	\centering
	
	\begin{tabular}{@{}c@{}}
		\includegraphics[width=\linewidth]{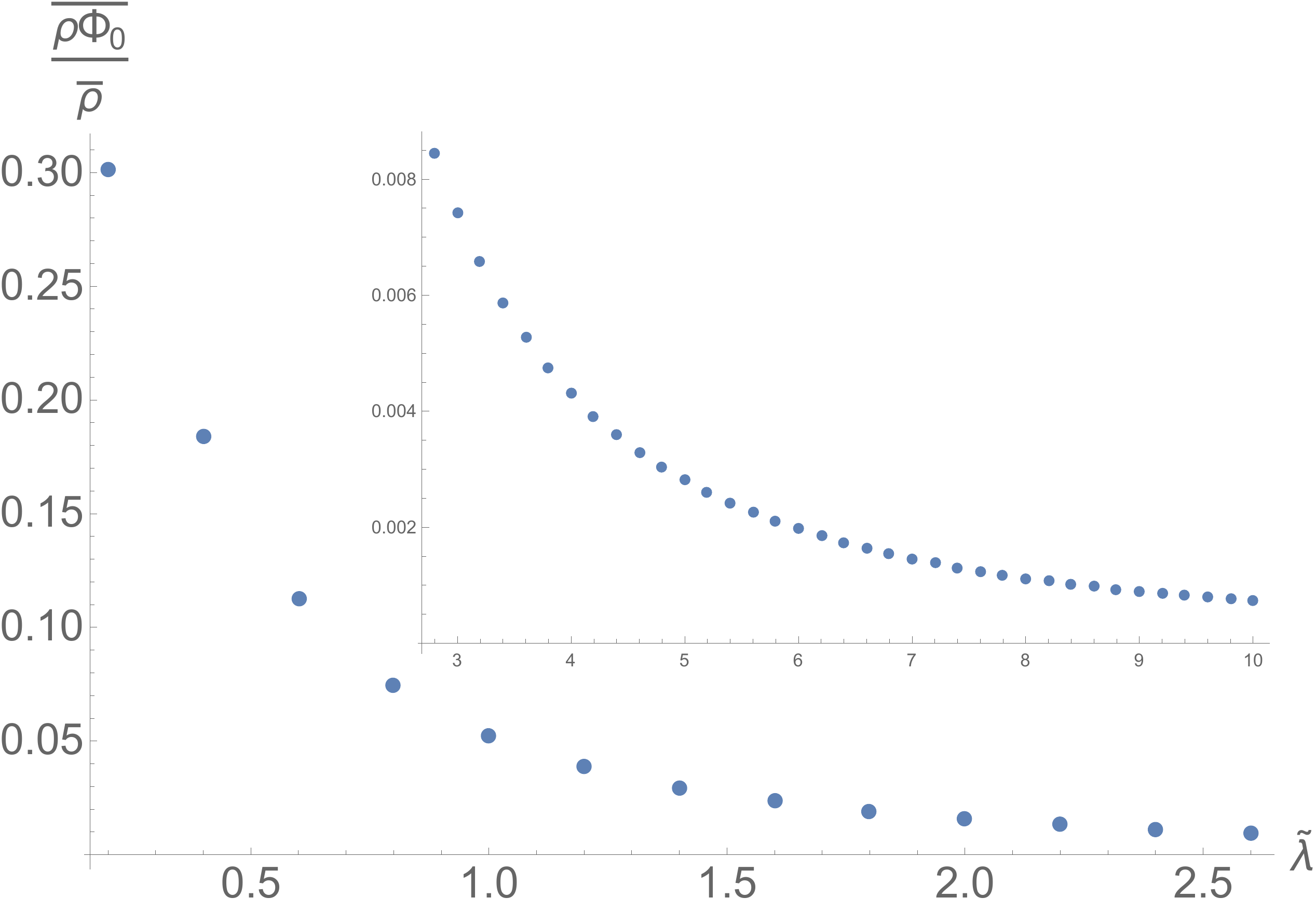} \\
	\end{tabular}
	
	
\caption{ Behavior of $\overline{\rho\Phi_0}/\overline{\rho}$  as a function of the renormalized screening length $\tilde\lambda$.
	}
	\label{fig.1}
\end{figure*}

Now, having these formulas, we can estimate numerically the quantities $\overline{\rho\Phi_0}/\overline{\rho}$ and  $\overline{\Phi^2_0}$. Figures \ref{fig.1} and 2 demonstrate behavior of
these quantities  as functions of $\tilde\lambda$.


\begin{figure*}[!ht]
	
	\vspace{1cm}
	
	\centering
	
	\begin{tabular}{@{}c@{}}
		\includegraphics[width=\linewidth]{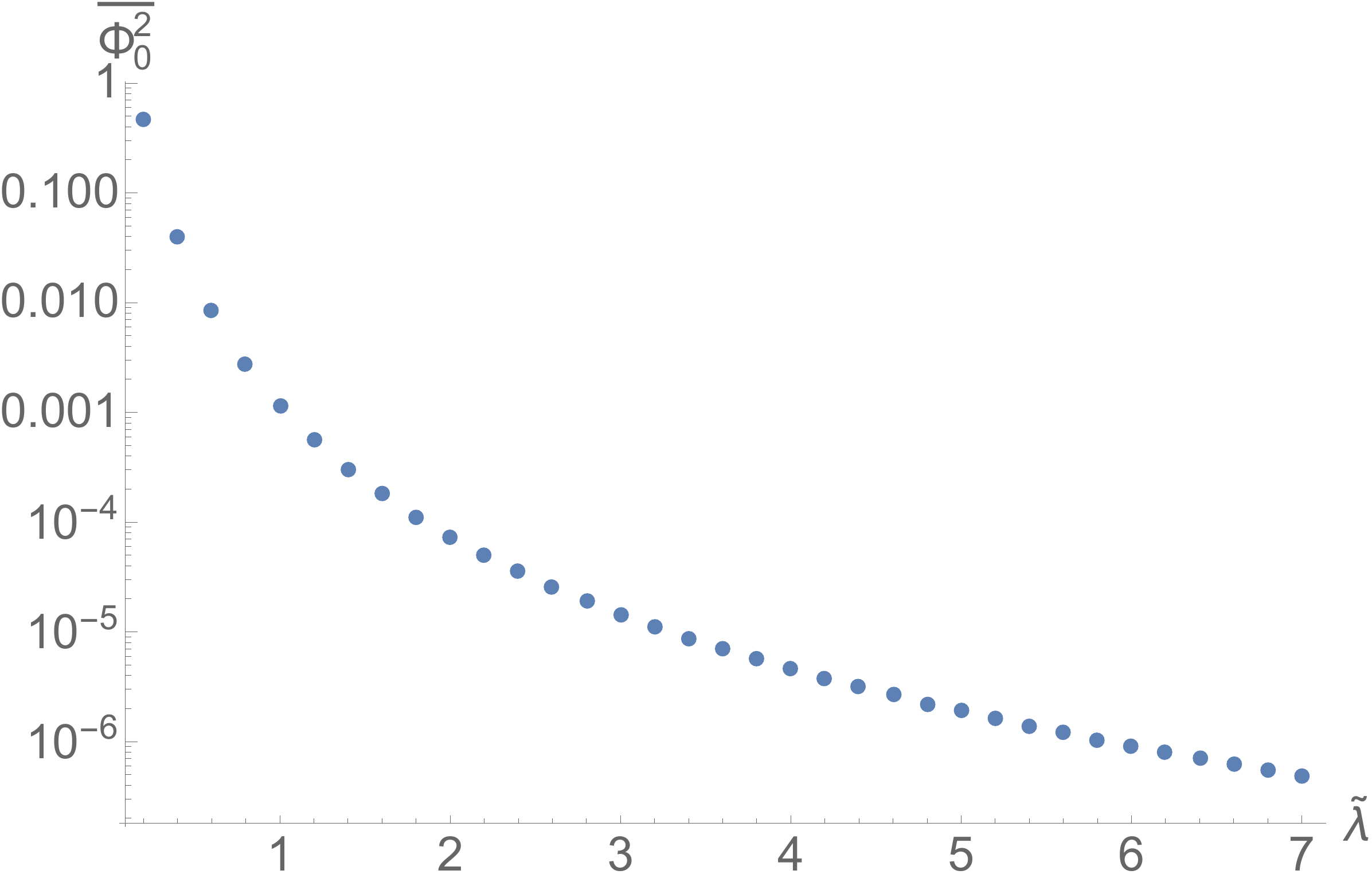} \\
	\end{tabular}
	
	\vspace{-2mm}
	
\caption{Behavior of $\overline{\Phi^2_0}$ as a function of the renormalized screening length $\tilde\lambda$.
	}
	\label{fig.2}
\end{figure*} 

\

In the case of the $\Lambda$CDM model, the renormalized screening length can be expressed as follows:
\be{2.19}
\tilde\lambda = \frac{\lambda}{a l} =\sqrt{\frac{2c^2}{9 H_0^2\Omega_{\mathrm{M}}}}\frac{1}{a_0 l}\frac{1}{\sqrt{z+1}} \approx \frac{3740\,\mathrm{Mpc}}{a_0 l} \frac{1}{\sqrt{z+1}}\, , 
\ee
where $a_0$ is the scale factor at the present time, $z$ is the redshift parameter, and for the illustrative purposes we took 
$H_0=67.4\,  \mathrm{km}\, \mathrm{s}^{-1} \mathrm{Mpc}^{-1}$
and $\Omega_{\mathrm{M}}=0.315$. This formula shows that the more the present day average distance between galaxies $a_0 l$ and the redshift $z$, the less the value of the renormalized screening length $\tilde\lambda$. Additionally, we see that both $\overline{\rho\Phi_0}/\overline{\rho}$ and $\overline{\Phi^2_0}$ are decreasing functions of  $\tilde\lambda$.
To get an idea of the numerical values of these quantities,
we take for the present day average distance between galaxies the value $a_0 l = 20$  Mpc and start from the redshift $z=100$. For these parameters we get $\tilde\lambda\approx 18.6$. Therefore, as it follows from Figures \ref{fig.1} and \ref{fig.2}, for $z<100$ we obtain 
$\overline{\rho\Phi_0}/\overline{\rho} < 10^{-3}$ and $\overline{\Phi^2_0}< 10^{-6}$. For example, at the present time 
$\tilde\lambda\approx 187$, and these second-order quantities are many orders of magnitude less than these upper limit values. Hence,
the effective average energy density $\overline{\varepsilon}^{(\mathrm{II})}(\eta)$ and pressure $\overline{p}^{(\mathrm{II})}(\eta)$ have a negligible backreaction effect on the Friedmann equations.


\section{Evaluation of ${\Psi_0^{(2)}}$}

Now, we evaluate the value of ${\Psi_0^{(2)}}$ which is the velocity-independent part of the second-order correction to the gravitational potential. According to Eq.~(3.10) in \cite{second-order2},
\ba{3.20}
\Psi_0^{(2)}&=&-\frac{3}{4}\Phi_0^2+\frac{\Phi_0}{6}-\frac{\pi \overline{\rho} \lambda}{ a}
\left(\frac{\kappa c^2}{8\pi a}\right)^2\sum\limits_{k}m_k\ e^{-a|\text{\textbf{r}}-\text{\textbf{r}}_k|/\lambda}\nonumber \nn\\
&+&\frac{1}{2}\left(\frac{\kappa c^2}{8\pi a}\right)^2{\sum\limits_{k,k'}}^{'}m_k m_{k'}\ \frac{
	e^{-a|\text{\textbf{r}}-\text{\textbf{r}}_{k}|/\lambda}}{|\text{\textbf{r}}-\text{\textbf{r}}_{k}|}
\frac{e^{-a|\text{\textbf{r}}_{k'}-\text{\textbf{r}}_k|/\lambda}}{|\text{\textbf{r}}_{k'}-\text{\textbf{r}}_k|}\, ,
\ea
where the prime denotes that $\text{\textbf{r}}_{k'} \neq \text{\textbf{r}}_k$. Obviously, in the case of identical particles of mass $m$ located at the same comoving distances $l$ from each other, the second sum for each $k$-th particle is the same. So, we can rewrite Eq. \rf{3.20} as follows:
\ba{3.21}
\Psi_0^{(2)}&=&-\frac{3}{4}\Phi_0^2+\frac{\Phi_0}{6}-\frac{\pi \overline{\rho} \lambda}{ a}
\left(\frac{\kappa c^2}{8\pi a}\right)^2m\sum\limits_{k}\ e^{-a|\text{\textbf{r}}-\text{\textbf{r}}_k|/\lambda} \nn \\
&+&\frac{1}{2}\left(\frac{\kappa c^2}{8\pi a}\right)^2m^2{\sum\limits_{k}}\ \frac{
	e^{-a|\text{\textbf{r}}-\text{\textbf{r}}_{k}|/\lambda}}{|\text{\textbf{r}}-\text{\textbf{r}}_{k}|}
{\sum\limits_{q}}^{'}\frac{e^{-ar_{q}/\lambda}}{r_{q}}  \, ,
\ea
where $r_{q}=l\sqrt{q_1^2+q_2^2+q_3^2}\neq0$. Introducing $\tilde{\bf r}={\bf r}/l$, $\tilde{\bf r}_k={\bf r}_k/l$, $\tilde\lambda=\lambda/(al)$ and taking into account that $m=\overline\rho l^3$, we further derive 
\ba{3.22}
\Psi_0^{(2)}&=&-\frac{3}{4}\Phi^2+\frac{\Phi}{6}-\pi\tilde\lambda
\left(\frac{\kappa c^2 m}{8\pi al}\right)^2\sum\limits_{k}\ e^{-|\tilde{\bf r}-\tilde{\bf r}_k|/\tilde\lambda}\nonumber \nn\\
&+&\frac{1}{2}\left(\frac{\kappa c^2m}{8\pi al}\right)^2{\sum\limits_{k}}\ \frac{
	e^{-|\tilde{\bf r}-\tilde {\bf r}_{k}|/\tilde\lambda}}{|\tilde{\bf r}-\tilde {\bf r}_{k}|}
{\sum\limits_{q}}^{'}\frac{e^{-\tilde r_{q}/\tilde\lambda}}{\tilde r_{q}}\, , 
\ea
where $\tilde r_{q}=\sqrt{q_1^2+q_2^2+q_3^2}\neq0$. In the case of our toy model, Eq. \rf{2.9} reads
\be{3.23} 
\Phi_0=\frac{1}{3}-\frac{\kappa c^2 m}{8\pi al}{\sum\limits_{k}}\ \frac{
	e^{-|\tilde{\bf r}-\tilde {\bf r}_{k}|/\tilde\lambda}}{|\tilde{\bf r}-\tilde {\bf r}_{k}|}\, ,
\ee
and from Eq. \rf{2.10} we also get $\kappa c^2 m/(8\pi al)=
a^2l^2/(12\pi \lambda^2)=1/(12\pi \tilde\lambda^2)$. Therefore,
substituting \rf{3.23} into \rf{3.22}, we get
\ba{3.24}
\Psi_0^{(2)}&=&-\frac{3}{4}\Phi_0^2+\frac{\Phi_0}{6}-\pi\tilde\lambda
\left(\frac{1}{12\pi\tilde\lambda^2}\right)^2\sum\limits_{k}\ e^{-|\tilde{\bf r}-\tilde{\bf r}_k|/\tilde\lambda}\nonumber \nn\\
&+&\frac{1}{2}\left(\frac{1}{12\pi\tilde\lambda^2}\right)\left(\frac{1}{3}-\Phi_0\right)
{\sum\limits_{q}}^{'}\frac{e^{-\tilde r_{q}/\tilde\lambda}}{\tilde r_{q}}\,  . 
\ea

According to the estimates in the previous section, the characteristic values of $\tilde \lambda$ in our Universe for $z\leq 100$ are much greater than 1. In this case, the sums in Eq. \rf{3.24} can be replaced by integrals:
\be{3.25} 
{\sum\limits_{q}}^{'}\frac{e^{-\tilde r_{q}/\tilde\lambda}}{\tilde r_{q}}\quad \Rightarrow\quad \int\limits_0^{+\infty}d\xi\frac{e^{-\xi/\tilde\lambda}}{\xi}4\pi\xi^2=4\pi\tilde\lambda^2\, 
\ee
and
\be{3.26} 
\sum\limits_{k}\ e^{-|\tilde{\bf r}-\tilde{\bf r}_k|/\tilde\lambda}\quad \Rightarrow\quad \int\limits_0^{+\infty}d\xi e^{-\xi/\tilde\lambda}4\pi\xi^2=8\pi\tilde\lambda^3\, ,
\ee
then, finally
\ba{3.27}
\Psi_0^{(2)}\; & \Rightarrow& \; -\frac{3}{4}\Phi_0^2+\frac{\Phi_0}{6}-\pi\tilde\lambda
\left(\frac{1}{12\pi\tilde\lambda^2}\right)^2 8\pi\tilde\lambda^3+\frac{1}{2}\left(\frac{1}{12\pi\tilde\lambda^2}\right)\left(\frac{1}{3}-\Phi_0\right)4\pi\tilde\lambda^2\nn \\
&=&-\frac{3}{4}\Phi_0^2\, . \ea
Since $|\Phi_0|\ll 1$, we obtain that $|\Psi_0^{(2)}| \ll |\Phi_0|$. Therefore, the second-order correction $\Psi_0^{(2)}$ is much less than the first-order quantity $\Phi_0$ as it should be.


\section{Conclusion}

In the present paper we have investigated the backreaction of  nonlinear perturbations on the global evolution of the Universe.  To this end, we have considered the second-order velocity-independent scalar perturbations within the cosmic screening approach. A remarkable feature of this approach is that nonlinear perturbations are presented in the analytical form. Thus, these perturbations can be investigated by analytical methods. 
For this purpose, we have considered the toy model in which all inhomogeneities (e.g., galaxies) have the same masses and are located at the same distance from each other. This allowed us to greatly simplify the original expressions. As a result, the numerical evaluation of the final expressions showed that nonlinear perturbations have a negligible backreaction effect on the Friedmann equations in agreement with the results obtained in \cite{Adamek2019}. We have also demonstrated that the second-order correction to the gravitational potential is much less than the corresponding first-order quantity. Consequently, the expansion of perturbations into orders of smallness in the cosmic screening approach is correct.


\section*{Acknowledgments}

The work of M. Eingorn, B. O'Briant and A. Diouf was supported by National Science Foundation (HRD Award \#1954454).


\end{document}